\documentstyle[preprint,aps]{revtex}
\begin{document}
\title{LANDAU SINGULARITY AND THE INSTABILITY OF VACUUM STATE IN QED}
\author{MOFAZZAL AZAM}
\address{THEORETICAL PHYSICS DIVISION, CENTRAL COMPLEX,
         BHABHA ATOMIC RESEARCH CENTRE\\
         TROMBAY,MUMBAI-400085,INDIA}
\maketitle
\begin{abstract}
In this paper we argue that the Landau singularity of the
coupling constant in QED reflects the instability of the
vacuum state, in other words the perturbative ground state of QED.
We do not have a rigorous mathematical proof. We rather provide
strong arguments derived from the theory of stability of matter
developed by Lieb {\it et al} on the one hand and the theory of
weakly interacting fermions developed by Feldman {\it et al}
on the other hand.

\end{abstract}
\vskip .8 in
\par Landau singularity of the coupling constant in
Quantum Electrodynamics has been known for more than
four decades \cite{landau}. The original derivation of the singularity
was based on the summation of one loop diagrams of vacuum
polarization tensor for photons in the perturbation theory.
In the early days the validity of such an expasion was looked upon
sceptically by many people including the authours  themselves.
However, after the advent of $1/N$ expansion technique, it was soon
realised that this singularity appears at the leading order in
$1/N_{f}$ expansion where $N_{f}$ is the number flavour or
number of specieses of electrons.This implies that in the
infinite flavour limit this singularity is exact.There have
been attempts to interpret this singularity in many ways.
Landau and Pomeranchuk tried  to argue that this singularity
reflects the fact that at short distances strong vacuum polarization
effects screen the electric charge completely.Others have called it
Landau ghost reflecting the internal inconsistency of
Quantum Electrodynamics \cite{shirkov}.In this paper, we shall argue that
Landau singularity reflects the instability of vacuum, in
other words the perturbative ground state of QED.We do not
have a rigorous mathematical proof. We rather rely on 'analogy
and precedence', the only other method available to a theoretical
physicist in the absence of a mathematical proof.For our 'analogy
and precedence' we refer to the theory of stability of matter
developed by Lieb {\it et al} on the one hand and the theory of
weakly interacting fermions developed by Feldman {\it et al}
on the other hand.

\par In quantum theory  a many-body system with ground
state energy $E_0$ is called thermodynamically stable
or simply stable if $ ~E_0/N ~$ is bounded bellow
when the number of particles, $~N\rightarrow \infty$.
In this context, the question of
stability of matter consisting of negatively charged electrons
and positively charged nuclei is very important.It was in 1967 when
Dyson and Lenard proved  that, in the framework of nonrelativitic
quantum mechanics, matter consisting of N electrons and K static
point nuclei is stable \cite{dyson}.
Subsequently, Lieb and his collaborators have made a very detailed
investigation of the stability of matter in nonrelativistic as well as
relativistic case ( \cite{lieb3,lieb1} and references there in ).
These studies seem to suggest that at
high enough energies Quantum Eletrodynamics may not be a well
behaved theory. Here we argue that there exists one approximation
scheme for QED in which it is really so - the perturbative ground
state of QED is unstable. This is the large flavour
limit of QED, where number of flavours $ ~N_{f} ~$ tends to infinity,
while $ ~e^{2}N_{f} ~$ is held fixed , "$e$" is the QED coupling constant.
Thermodynamic stability of a system of particles interacting via
electromagnetic interaction is associated with the control of the
short distance behaviour of the interaction. In the nonrelativistic
limit, zero point kinetic energy of the fermions controls this short
distance behavoiur- however, in the relativistic limit, there is
a need for certain bounds on the value of the fine structure constant
\cite{lieb3,lieb1,conlon}.
In the large flavour limit of QED, Landau singularity becomes exact
and it is well known that this is  also associated with the short distance
behaviour of the electromagnetic interaction \cite{landau} .
We refer to the analogy of this model with the theory of
weakly interacting electron system analysed
by Feldman {\it et al} using renormalisation group technique.We find an
exact analogy between the Landau singularity in QED and
the singularity of effective coupling constant in some angular
mometum channel in the condensed matter system.In the condensed
matter system, we know that the singularity reflects
the instability of Landau-Fermi liquid
ground (generally called the BCS instability).
This analogy clearly suggests that the Landau singularity
reflects the instability of ground state in QED.
We shall argue in this paper using the example
of large flavour QED, that the
singularity of effective running coupling constant in the second quantized
quantum theory is the equivalent of thermodynamic
instability in many-body systems of particles in the first
quantized quantum theory.
\par It should be mentioned here that the stability of ground state
and the possible existence an ultra-violet fixed point has been studied
extensively in the lattice formulation of QED in the wide range of values
of the fermion flavour $N_f$ by Kogut {\it et al} \cite{kogut1,kogut2}.
On the other hand in
massless QED in the continuum, Miransky  argues that there exists a
chiral symmetry breaking phase \cite{mir} .However,
in these studies Landau
singularity plays no role.There has been some recent studies of Landau
singularity using lattice formulation of QED by Gockeler {\it et al}
\cite{gock}.
These studies seem to suggest that chiral symmetry breaking allows
QED to escape Landau singularity.But then chiral symmetry breaking,
as their
study shows, seems to be intimately connected with trivility of QED.
Our approach is motivated by the problems in the theory of stability
of matter as the following section would elaborate.We are interested
in finding the meaning of Landau singularity rather than finding a way
to escape it. (However, we do believe that it is important to investigate
the consequences of escaping the singularity as the study in
\cite{gock} seem to suggest)
In our approach Landau singularity is not treated as a ghost
as in many other publications but as a signature of instability of the
underlying ground state.

\section{The Large Flavour QED}
\par The stability of many-body system of fermions corresponding to
large flavour QED can be considered in two different ways.One way is to
assume that at lower energies only a few flavours are excited.
In that case it
becomes similar to the single flavour case.
In this way we avoid confusion that may arise
due to the double limit, number of particles $N$ and number flavour
$N_{f}$, both tending to infinity.
At high energies when we need to consider  QED,
all specieses of fermions
are excited and to avoid additional complication , they can be
assumed to be of the same mass.However, we believe this is not the
correct way- here we miss the role played by flavour in a many-body
theory.
It is possible to consider the
quantum mechanical many-body problem with large flavours by
carefully taking care of the double limit mentioned above.
Let us consider a region of linear dimension $R$ where there are $N$
number of fermions with $N_f$ flavours.Assume $N>>N_f$.Taking into
account the the Pauli exclusion principle for the case
involoving multiflavour fermions in $3$-dimensional space,
the kinetic energy K can be written \cite{yau}
as $~~K \approx \frac{\hbar^2}{2m} \frac{N^{5/3}}{R^{2}N_{f}^{2/3}}$.
The potential energy $U$ due to Coulomb interaction
is given by \cite{lieb3},
$~~U=-a_0 \frac{e^{2}N^{4/3}}{R}$,$~~a_0>0~~$ is a  numerical constant.
The total energy, $E(R)$, is given by
\begin{eqnarray}
E(R)=K+U \approx \frac{\hbar^2}{2m} \frac{N^{5/3}}{R^{2}N_{f}^{2/3}}
-a_0 \frac{e^{2}N^{4/3}}{R}
\end{eqnarray}
The ground state is obtained by minimizing the total energy,
$E(R)$, with respect to $R$.
\begin{eqnarray}
E_0~~\approx~ -\frac{ma_{0}^{2}}{\hbar^{2}}(e^2 N_f^{1/3})^2 N ~;~~~
\epsilon_0 ~=~E_0/N \approx~ -\frac{ma_{0}^{2}}
{\hbar^{2}}(e^2 N_f^{1/3})^2
\nonumber\\
R_0~~\approx~\frac{\hbar^2}{ma_0}
(\frac{N}{N_f})^{1/3}\frac{1}{e^2 N_f^{1/3}}~;~~~
r_0=\frac{R_0}{(N/N_f)^{1/3}} \approx
~\frac{\hbar^2}{ma_0}\frac{1}{e^2 N_f^{1/3}}
\end{eqnarray}
When the number of flavour $N_f$ increases, the size of the small box
$r_0$ decreases and when it approaches Compton wave length
$\frac{\hbar}{mc}$, we can no longer use the nonrelativistic quantum
mechanics.Therefore, the above estimates are correct only when
\begin{eqnarray}
N_f~<<~ \frac{1}{a_{0}^3} \frac{1}{(\frac{e^2}{\hbar c})^3}
\end{eqnarray}
In the relativistic quantum theory of many-body systems, it is
almost customary now to take $~c|p|~$ \cite{lieb3},
where $c$ is the velocity of light,
as the kinetic energy of
individual particles.Therefore, the kinetic energy $K$ for the system
of $N$ particles with $N_f$ flavour in a region with linear dimension
$R$ is given by, $K~=~\frac{\hbar c N^{4/3}}{R N_{f}^{1/3}}~~$.
Therefore,in the relativistic case,
\begin{eqnarray}
E(R)=~K~+~U~\approx~\frac{\hbar c N^{4/3}}{R N_{f}^{1/3}}~
-a_0 \frac{e^{2}N^{4/3}}{R}
\end{eqnarray}
In this case stability of ground state requires $E(R)~>~ or~=0$.This
leads to the condition,
\begin{eqnarray}
N_f~<~ or~~=\frac{1}{a_{0}^3} \frac{1}{(\frac{e^2}{\hbar c})^3}
\end{eqnarray}
which in the unit $~c~=\hbar~=1~$, becomes
$N_f~<~ or~~=\frac{1}{(a_{0}e^{2})^3}~$.
At this ponit a few comments are necessary.We considered a theory
with large $N$ and large $N_f$, and it is implicit in our discussion
that we are considering, $N~\rightarrow~\infty$, but we have said
nothing
about the flavour $N_f$.We could have also taken
$N_f~\rightarrow~ \infty$ but slower than N. We could also assume that
in the limit $N_f~\rightarrow~\infty$ and $e~\rightarrow~0~~$,
$e^{2} N_{f}^{1/3}~=constant$, and satisfies the equation Eq.[5]
and conclude that in
this limit matter is stable.
Mathematically it is correct.But this is only  a formal result.$e$  does
not tend to zero and $N_f$ , never tends to infinity.
The mathematical limit, has the
physical implication that when $e$ is small and $N_f$ is large,the
relevant parameter that sets the stability criteria
is $e^2 N_{f}^{1/3}$.It is the value of
$e^2 N_{f}^{1/3}$   that decides the stability of the system.When
$e^2 N_{f}^{1/3}~>~\frac{1}{a_{0}^{3}}~$, in other words, when
the number of flavours
$N_f~>~ \frac{1}{(a_{0}e^{2})^3}~$,
the {\bf many-body system is thermodynamically unstable}.
\par In Quantum
Electrodynamics we find that the relevant parameter is not
$e^2 N_{f}^{1/3}$ but $e^2 N_{f}$.
In large flavour QED with $ ~N_{f} ~$ flavours, when we calculate
the connected Green's functions using Feynman diagram technique,
we find that one loop diagram are of order $e^2 N_{f}$ and higher
loop diagrams are of lower order in $N_f$.This becomes
more transparent by
carrying out expansion in
$1/N_{f}$.
The essential idea is
to replace $e{\rightarrow}e/\sqrt{N_{f}}$, and consider
the Lagrangian,
\begin{equation}
{\cal L} =  \sum_{i=1}^{N_f}~ \bar{{\Psi}}^i
\Big(i\gamma^{\mu} \partial_{\mu}
+ m - e/\sqrt N_f ~\gamma^{\mu} A_{\mu}\Big) \Psi^{i}
+\frac{1}{4} F_{{\mu}{\nu}}^2
\end{equation}
where $ \psi$  is   the    Dirac    field,    and   $ A_{\mu}$    and
$F_{{\mu}{\nu}}$  are  the
potential  and the field strength respectively.
In the leading (zeroth) order in $ ~1/N_f ~$ ,
the photon propagator receives radiative corrections only
from the the single-loop Feynman diagrams.Higher loop contributions
are of order one or more in $1/N_{f}$.
The renormalized running (effective) coupling constant at energy,
$ ~k^2={\Lambda}^2 ~ $, up to the leading order in $ ~1/N_f ~$ ,
is given by
\begin{equation}
e_{eff}^{2}(\Lambda^2) = {e^2  \over 1- (e^2/3\pi) ln \Lambda^2 /m^2}
\end{equation}
where we have used, $e^2=e^2(k^2=m^2)$, for the physical coupling
constant.
One  can  easily infer that the running coupling constant blows up at,
$ ~\Lambda^2 = m^2  exp( 3\pi /e^2) ~$.
This is known as {\bf Landau singularity} in QED \cite{landau}.
It is easy to see that this
singularity is associated with the short distance (large momentum)
behaviour of the interaction.
To understand the meaning of Landau singularity,
here we refer to the renormalisation group analysis of weakly
interacting Fermi system with short range potential at
finite density and zero
temparature by Feldman {\it et al}
\cite{feld,froh}.
The iterative renormalization group transformations
leads to the same type of singularity in
some  suitably defined running coupling constant.

\section{Feldman Model of Weakly Interacting Electron Gas}
We briefly describe here the Feldman Model of weakly interacting
electron gas (see reference \cite{froh} for details).
The model of weakly interacting electron gas studied by
Feldman {\it et al}
is a condensed matter Fermi system in thermal equillibrium
at some temparature $T$ (for simplicity, assume $T=0$)
and chemical potential $\mu$.
On microscopic
scale($\approx 10^{-8}$ cm), it can be described approximately
in terms of non-relativistic electrons with
short range two body interactions.
The thermodynamic quantities such as conductivity depend only on
physical properties of the system at mesoscopic length scale
($\approx 10^{-4}$ cm),
and therefore, are determined from processes involving momenta
of the order of
$\frac{k_{F}}{\lambda}$ around the Fermi surface, where
the parameter, $\lambda >>1$,  should be thought of
as a ratio of meso-to-microscopic length scale. This is
generally refered to as
the scaling limit(large $\lambda$, low frequencies) of the system.
The most important observation of Feldman et. al.
is that in the scaling
limit, systems of non-relativistic (free)
electrons in $d$ spatial dimensions behave like a system of
multi-flavoured relativistic chiral Dirac fermions
in $1+1$ dimensions.The number of flavours
$N \approx ~const. ~ \lambda^{d-1}$.
It is possible then to set up a renormalization group
improved perturbation
theory in $\frac{1}{\lambda}$ around the non-interacting electron gas,
where in, the large number of flavours $N$, play an important role
in actual calculations.
\subsection{Free Electron gas and the Multiflavour Relativistic
Fermions in $1+1$ Dimensions}
Let us consider a system of non-relativistic
free electrons in $d$ spatial dimensions with the Euclidean action,

\begin{eqnarray}
S_{0}(\psi^*,\psi)=\sum_{\sigma} \int d^{d+1} x \psi_{\sigma}^*(x)
(\rm{i} \partial_0-\frac{1}{2m} \Delta -\mu)\psi_{\sigma}(x)
\end{eqnarray}
The Euclidean free fermion Green's function, $ G^{0}_{\sigma \sigma'}
(x-y)$,
where $\sigma$ and $\sigma'$  are the spin indices,
$x=(t,\vec{x})$ and $y=(s,\vec{y})$,
$t$ and $s$ are imaginary times,
$t>s$, is given by,

\begin{eqnarray}
G^{0}_{\sigma \sigma'}(x-y)=\langle \psi_{\sigma}^*(x)\psi_{\sigma}(y)
\rangle_{\mu}
=-\delta_{\sigma \sigma'} \int (dk)\frac{e^{-ik_0(t-s)+i\vec{k}
(\vec{x}- \vec{y})}}{ik_0 -(\frac{k^2}{2m}-\mu)}
\end{eqnarray}
Where we have used $(dk)=\frac{1}{(2\pi)^(d+1)} d^{d+1}k$.
In the scaling limit, the leading contributions to
$ G^{0}_{\sigma \sigma'}(x-y)$
come from modes whose momenta are contained in a
shell $S_{F}^{(\lambda)}$ of thickness $\frac{k_{F}}{\lambda}$
around the Fermi surface $S_{F}$.
In order to approximate the Green's function, let us
introduce the new variables
$\vec{\omega}, ~p_{\parallel}, ~p_{0}$ such that $k_{F}\vec{\omega}
\in S_F$, $p_0 =k_0$ and $\vec{k}=(k_F+p_{\parallel})\vec{\omega}$.
If $\vec{k}\in S_{F}^{(\lambda)}$, then $p_{\parallel}<<k_{F}$, and
we can approximate the integrand of Eq.(4), by dropping
$p_{\parallel}^{2}$ term in the denominator.In other words,
\begin{eqnarray}
G^{0}_{\sigma \sigma'}(x-y)= \delta_{\sigma \sigma'}\int
\frac{d {\bf \sigma} (\vec{\omega})}{(2\pi)^{d-1}} k_{F}^{d-1} e^{ik_{F}
\vec{\omega}(\vec{x}-\vec{y})} G_{c}(t-s, \vec{\omega}(\vec{x}-\vec{y})
\end{eqnarray}
where $d {\bf \sigma}(\vec{\omega})$ is the uniform measure on unit
sphere and
\begin{eqnarray}
G_{c}(t-s, \vec{\omega}(\vec{x}-\vec{y}))=-\int \frac{dp_{0}}{2\pi}
\frac{dp_{\parallel}}{2\pi} \frac{e^{-i k_0 (t-s)+ip_{\parallel}
\vec{\omega}(\vec{x}- \vec{y})}}{ip_0 -v_{F}p_{\parallel}}
\end{eqnarray}
is the Green's function of chiral Dirac fermion in $1+1$ dimension.
$v_{F}=k_{F}/m$ is the  Fermi velocity.The $\vec{\omega}$-integration
in Eq.(5) can be further approximated by replacing it with summation
over discrete directions $\vec{\omega}_{j}$ by dividing the shell
$S_{F}^{(\lambda)}$ into $N$ small boxes $B_{\vec{\omega}_{j}}, j=1,..,N$
of roughly cubical shape.The box, $B_{\vec{\omega}_{j}}$, is centered at
$\vec{\omega}_{j} \in S_F$ and has an approximate side length
$\frac{k_F}{\lambda}$.The number of boxes, $N=\Omega_{d-1} \lambda ^{d-1}$,
where $\Omega_{d-1}$ is the surface volume of unit sphere in
$d$ spatial dimensions.The Green's function is, now, given by
\begin{eqnarray}
G^{0}_{\sigma \sigma'}(x-y)=-\delta_{\sigma \sigma'}\sum_{\vec{\omega}_{j}}
\int \frac{dp_{0}}{2\pi} \frac{dp_{\parallel}}{2\pi} \frac{p_{\perp}}
{2\pi}
\frac{e^{-ip_0(t-s)+i\vec{p} (\vec{x}- \vec{y})}}
{ip_0 -v_{F}p_{\parallel}}
\end{eqnarray}
where $\vec{p}=p_{\parallel}\vec{\omega}+\vec{p}_{\perp}$ is a vector
in $B_{\vec{\omega}_{j}}-k_{F} \vec{\omega}_{j}$ and $p_{0}\in
\mathcal{R}$.
Thus in the scaling limit, the behaviour of a $d$-dimensional
non-relastivistic free electron gas is described by
$N=\Omega_{d-1} \lambda ^{d-1}$ flavours of free chiral Dirac fermions
in $1+1$ dimensional space-time.The propagator
$G_{c}(t-s, \vec{\omega}(\vec{x}-\vec{y})$ depends on the flavour
index $ ~\vec{\omega} ~$.But the energy of an electron or hole with momenta
$ ~\vec{k} ~$ , depends only on $p_{\parallel}$, where
$ ~p_{\parallel}=\vec{k}.\vec{\omega}-k_F ~$ and
$ ~\vec{\omega}=\frac{\vec{k}}{|k|} ~$.It is proportional to
$ ~p_{\parallel} ~$ , just as for relativistic fermions in
$1+1$ dimensions.
\subsection{Renormalization Group Flow and the BCS Instability}
Large scale behaviour of the weakly interacting system is
described by an effective action.
To see how the effective action is calculated,
let us consider a system with Euclidean action of the form,
\begin{eqnarray}
S(\psi^*,\psi)=S_{0}(\psi^*,\psi) + S_{I}(\psi^*,\psi)
\end{eqnarray}
where $S_{0}(\psi^*,\psi)$ is the quadratic part given in Eq.(3)
and $S_{I}(\psi^*,\psi)$ is the quartic interaction term.For a weakly
interacting electron gas,
\begin{eqnarray}
S_{I}(\psi^*,\psi)=g_{0}k_{F}^{1-d}\sum_{\sigma,\sigma'}\int d^{d+1}x
d^{d+1}y :\psi_{\sigma}^{*}(x) \psi_{\sigma}(x) v(\vec{x}-\vec{y})
\delta(x_0-y_0) \psi_{\sigma'}^{*}(y) \psi_{\sigma'}(y)
\end{eqnarray}
The factor $k_{F}^{1-d}$ ensures that $g_{0}$ is
dimensionless. The two body potential $v(\vec{x})$ is
assumed to be smooth and short range, and therefore, its Fourier
transform $\hat{v}(\vec{k})$ is also smooth.
In order to calculate the effective actions,
we first split the field variable into  slow and fast modes,
$ ~\hat{\psi}=\hat{\psi}_{<}+\hat{\psi}_{>} ~$, where
Supp $\hat{\psi}_{>} \subset R \times
(R^d ~ \setminus ~
S_{F}^{(\lambda)}) ~~(\rm{region} >) ~~$ and
Supp $\hat{\psi}_{<} \subset R \times S_{F}^{(\lambda)}
 ~~(\rm{region} <) ~~$.
We then integrate out the fast modes using the functional integral
for fermions,
\begin{eqnarray}
e^{-S_{eff}(\hat{\psi}^{*}_{<},\hat{\psi}_{<})}=
\frac{1}{\Xi}\int D\hat{\psi}^{*}_{>} D\hat{\psi}_{>}
e^{-S(\hat{\psi}^{*}_{>},\hat{\psi}^{*}_{<},\hat{\psi}_{>},\hat{\psi}_{<})}
\end{eqnarray}
Now using the linked cluster theorem, we obtain
\begin{eqnarray}
e^{-S_{eff}(\hat{\psi}^{*}_{<},\hat{\psi}_{<})}=
\exp{(-S_{0,<}-<S_{I}>_{G^{0}_{>}}+\frac{1}{2}<S_{I};S_{I}>_{G^{0}_{>}}
-\frac{1}{3}<S_{I};S_{I};S_{I}>_{G^{0}_{>}}-...)}
\end{eqnarray}
The abbreviations, $<a;b>$ $<a;b;c>$ etc., denote
the connected correlators.
The subscripts $G^{0}_{>}$ indicate that the
expectations $<(.)>_{G^{0}_{>}}$ are calculated using infrared cut-off
free propagators  in accordance with the functional measure,
$dP(\hat{\psi}^{*}_{>},\hat{\psi}_{>})
=(1/\Xi) D\hat{\psi}^{*}_{>} D\hat{\psi}_{>}
e^{-S_{0}(\hat{\psi}^{*}_{>},\hat{\psi}_{>})}$.
The connected correlators can be evaluated using the Feynman
diagram technique. Therefore, the effective action can be calculated
once we know the amplitudes of connected Feynman diagrams.
It is clear that
the $S_{eff}$ contains far more interactions than the original
quartic interaction $S_{I}$.However, for weakly interacting systems
the original coupling remains dominant.
To carry out the iterative
renormalization group  scheme of Feldman {\it et al}, we
choose some large scale $\lambda_{0}<< \frac{1}{\sqrt{g_0}}$ and
calculate the effective action, $S_{eff}$ perturbatively to leading order
in $\frac{1}{\lambda_0}$.The effective action depends on modes
corresponding to wave vectors located in the shell, $S_{F}^{(\lambda)}$,
of width
$\frac{k_F}{\lambda _0}$ around the Fermi surface.
Now, we divide the shell
$S_{F}^{(\lambda)}$ into $N=const. ~ \lambda^{d-1}$ cubical
boxes, $B_{\vec{\omega}_{j}}$,
of approximate  side length
$\frac{k_F}{\lambda _0}$.Next, we rescale all the momenta so that,
instead of belonging to the boxes $B_{\vec{\omega}_{j}}$
they are contained in boxes
$\tilde{B}_{\vec{\omega}_{j}}$ of side length $\approx k_{F}$.
These two steps are generally known as decimation of degrees of
freedom and rescaling.
The renormalization group scheme
consists of iteration of these two steps.
\par Assume that the degrees of freedom corresponding to momenta
not lying in $S_{F}^{(\lambda)}$ have been integrated out.
Let $\hat{\psi}_{\sigma}(k), k\in R \times S_{F}^{(\lambda)}$,
denote these modes. The sector fields are defined as
\begin{eqnarray}
\psi_{\vec{\omega},\sigma}(x)=\int_{R \times B_{\vec{\omega}}}(dk)
e^{i(k_{0}t-(\vec{k}-k_{F}\vec{\omega})\vec{x})} \hat{\psi}_{\sigma}(k)
\end{eqnarray}
It is easy to see that
$ ~\psi_{\sigma}(x)=\sum_{\vec{\omega}} e^{\rm{i} k_{F}\vec{\omega}\vec{x}}
\hat{\psi}_{\vec{\omega},\sigma}(x) ~$.
Inserting the Fourier transform of sector fields in Eq.(3) and Eq.(9),
and carrying out some algebraic manipulations, we obtain
\begin{eqnarray}
S_{0}=-\sum_{\vec{\omega},\sigma}\int_
{\mathcal{R}\times (B_{\vec{\omega}}-k_{F}\vec{\omega})}(dp)
\hat{\psi}^{*} _{\vec{\omega},\sigma}(p)\big(\rm{i}p_{0}-v_{F}\vec{\omega}
\vec{p} +O\big(\frac{1}{\lambda^{2}}\big)\big)
\hat{\psi}_{\vec{\omega},\sigma}(p)
\end{eqnarray}
\begin{eqnarray}
S_{I}&=&\frac{g_{0}}{2} k_{F}^{1-d} \sum_{\vec{\omega}_1,..,\vec{\omega}_4
;\sigma,\sigma'}\hat{v}(k_F(\vec{\omega}_1- \vec{\omega}_4)) \int
(dp_{1})...(dp_{4})(2\pi)^{d+1} \delta (p_1+p_2- p_3-p_4)  \nonumber\\
& &\hat{\psi}^{*} _{\vec{\omega}_1, \sigma}(p_1)
\hat{\psi}^{*} _{\vec{\omega}_2,\sigma'}(p_2)
\hat{\psi}_{\vec{\omega}_3,\sigma'}(p_3)
\hat{\psi}_{\vec{\omega}_4,\sigma}(p_4) +
\rm{terms ~~of ~~higher ~~order
 ~~in}  \frac{1}{\lambda}
\end{eqnarray}
Using cluster expansions to integrate out the degrees of freedom
corresponding to
momenta outside the shell $S_{F}^{(\lambda_0)}$, one can show
that , at scale $\frac{k_F}{\lambda}$ , the effective action has the
form given by Eq.(13) and Eq.(14) except that
$\hat{v}(k_F(\vec{\omega}_1- \vec{\omega}_4))$ is replaced by
a coupling function
$g(\vec{\omega}_1,\vec{\omega}_2,\vec{\omega}_3,\vec{\omega}_4)
\approx \hat{v}(k_F(\vec{\omega}_1- \vec{\omega}_4))$ with
$\vec{\omega}_1 + \vec{\omega}_2 = \vec{\omega}_3 + \vec{\omega}_4$.
\par Next, one considers the rescaling of the fields and the action.The
fields are rescaled in such a that the supports of the Fourier
transformed "sector fields" are boxes,
$\tilde{B}_{\vec{\omega}}=\lambda (B_{\vec{\omega}}-k_{F}\vec{\omega})$
,of roughly cubical shape with
sides of length $k_{F}$, and such that the quadratic part of the action
remains unchanged to leading order in $\frac{1}{\lambda}$.The first
condition implies that
$p \longmapsto \tilde{p}=p\lambda$ and $x \longmapsto \xi=
\frac {x}{\lambda}$.
The rescaled
sector fields and their Fourier transforms are given by,
\begin{eqnarray}
\tilde{\psi}_{\vec{\omega},\sigma}(\xi)=\lambda^{\alpha}
\psi_{\vec{\omega},\sigma}(\lambda \xi) ~~~;~~~
\hat{\tilde{\psi}}_{\vec{\omega},\sigma}(\tilde{p})=\lambda^{\alpha-d-1}
\hat{\psi}_{\vec{\omega},\sigma}(\frac{\tilde{p}}{\lambda})
\end{eqnarray}
Inserting the scaled Fourier transformed fields into quadratic part of
action $S_{0}$, it is easy to see that $S_{0}$ remains unchanged if
the scaling dimension $\alpha=\frac{d}{2}$.
Now, inserting the rescaled fields in the quartic part of the action,
we find that the quartic part has scaling dimension $(1-d)$.
The quartic terms of higher degree in momenta as well as terms of
higher degree in fields appearing in the effective action have
smaller scaling dimensions.
\newpage
Thus the effective action in terms of scaled sector fields
is,
\begin{eqnarray}
S_{eff}&=&\sum_{\vec{\omega},\sigma}\int (d \tilde{p})
\hat{\tilde{\psi}}^{*}_{\vec{\omega},\sigma}(\tilde{p})
(\rm{i}\tilde{p}_{0}-v_{F}\vec{\omega}
\vec{\tilde{p}})
\hat{\tilde{\psi}}_{\vec{\omega},\sigma}(\tilde{p})
+\frac{1}{2} \frac{1}{\lambda^{d-1}}
\sum_{\vec{\omega}_1+\vec{\omega}_2=\vec{\omega}_3+\vec{\omega}_4
; \sigma,\sigma'}
g(\vec{\omega}_1,\vec{\omega}_2,\vec{\omega}_3,\vec{\omega}_4) \nonumber\\
& &\int
(d\tilde{p}_{1}). \ldots (d\tilde{p}_{4})(2\pi)^{d+1}
\delta (\tilde{p}_1+\tilde{p}_2- \tilde{p}_3-\tilde{p}_4)
\hat{\tilde{\psi}}^{*} _{\vec{\omega}_1, \sigma}(\tilde{p}_1)
\hat{\tilde{\psi}}^{*} _{\vec{\omega}_2,\sigma'}(\tilde{p}_2)
\hat{\tilde{\psi}}_{\vec{\omega}_3,\sigma'}(\tilde{p}_3)
\hat{\tilde{\psi}}_{\vec{\omega}_4,\sigma}(\tilde{p}_4) \nonumber\\
& &+\rm{terms ~~of ~~higher ~~order ~~in}  \frac{1}{\lambda}
\end{eqnarray}
We see that that the inverse propagator for the sector field is
diagonal in $\vec{\omega}$ ,and it depends only on $p_0$ and
$p_{\parallel} =\vec{\omega}\vec{p}$ but not on
$p_{\perp} =\vec{p}-(\vec{\omega}.\vec{p})\vec{\omega}$.
\par We are interested in the renormalization group
flow equations in the leading order in $\frac{1}{\lambda}$.
Therefore, for carrying out the decimation of degrees of freedom,
we will be interested only in those diagrams that contribute
to the amplitude in the leading order in $\frac{1}{\lambda}$.Before
we consider these diagrams, let us consider the possible inter sector
scattering geometries.How many independent
$g^{(0)}(\vec{\omega}_1,\vec{\omega}_2,\vec{\omega}_3,\vec{\omega}_4)$
exists?
For $d=3$, suppose $\vec{\omega}_3 \neq -\vec{\omega}_4$.On the unit
sphere, there  are $N^{(0)}=Const.\lambda_{0}^{d-1}$
different $\vec{\omega}$'s.
But all choises of $\vec{\omega}_1$ and $\vec{\omega}_2$ with
$\vec{\omega}_1 + \vec{\omega}_2 = \vec{\omega}_3 + \vec{\omega}_4$ lie
on a cone containing $\vec{\omega}_3$  and $\vec{\omega}_4$
with $\vec{\omega}_3 + \vec{\omega}_4$ as the symmetry axis.Therefore,
there are $O(\lambda_{0}^{d-2})$ choices.Only when
$\vec{\omega}_3 = - \vec{\omega}_4$ that there are
$N^{(0)}=Const.\lambda_{0}^{d-1}$ choices.
Couplings involving incoming states with
$\vec{\omega}_3 \neq -\vec{\omega}_4$ will be represented by
$g^{(0)}(\vec{\omega}_1,\vec{\omega}_2,\vec{\omega}_3,\vec{\omega}_4)$.
Couplings that involve sectors $\vec{\omega}_3 = -\vec{\omega}_4$
or equivalently $\vec{\omega}_1 = -\vec{\omega}_2$ will be
denoted by $g^{(0)}_{BCS}(\vec{\omega}_1,\vec{\omega}_4)$. Because of
rotational invariance, $g^{(0)}_{BCS}(\vec{\omega}_1,\vec{\omega}_4)$
is a function of only the angle between $\vec{\omega}_1$
and $\vec{\omega}_4$.
\par The chemical potential receives corrections
from connected diagrams with two external electron lines (self energy
correction of the electrons).
The contribution of order zero in
$\frac{1}{\lambda}$ comes only from the
{\bf tadpole} and {\bf turtle} diagrams.
These diagrams contain one internal interaction squiggle of order
$\frac{1}{\lambda_{0}^{d-1}}$ and there are
$N^{(0)}=Const.\lambda_{0}^{d-1}$
choices of the inner particle sector.Therefore, the contribution
is of order zero in $\frac{1}{\lambda}$. The amplitude corresponding
to these tadpole and turtle diagrams turns out to be $p$-independent but
of order $O(g^(0))$.Thus, $ ~\delta \mu_{1}= ~O(g^{(0)}/\lambda_{0}) ~$,
and the renormalized electron propagator is
$ ~G^{ ~R ~}_{\vec{\omega}}(p_{0}, \vec{p}.\vec{\omega}) ~
=- ~[\rm{i}p_{0}-v_{F}\vec{p}.\vec{\omega}+\lambda_{0}
\delta \mu_{1}]^{-1}$.
\par To find the evolution of the coupling constant
$g(\vec{\omega}_1,\vec{\omega}_2,\vec{\omega}_3,\vec{\omega}_4)$,
we have to calculate amplitude of diagrams with four external legs.
It is found that when
$\vec{\omega}_1 + \vec{\omega}_2 =\vec{\omega}_3 + \vec{\omega}_4 \neq 0$,
the coupling functions
do not flow in the leading order in $\frac{1}{\lambda}$.
But for sector indices
$\vec{\omega}_1 + \vec{\omega}_2 =\vec{\omega}_3 + \vec{\omega}_4 = 0$,
the coupling functions, $g^{(0)}_{BCS}(\vec{\omega}_1,\vec{\omega}_4)$,
flow.
The diagrams that contribute to the flow equation
are the {\bf ladder diagrams} with
self energy insertion for the internal electron lines but with no other
two legged subdiagram.The amplitude of such a diagram with
$n$ interaction squiggles and with zero incoming and outgoing
box momenta of the particles is given by,
\begin{eqnarray}
\big( \frac{1}{\lambda^{d-1}} \big)^{n+1}
\sum_{\vec{\omega_1},...,\vec{\omega_n}}
(-1)^n \beta^n g_{BCS}(\vec{\omega},\vec{\omega}_n)
g_{BCS}(\vec{\omega}_n,\vec{\omega}_{n-1}) ...
g_{BCS}(\vec{\omega}_1,\vec{\omega}^{\prime})  \nonumber
\end{eqnarray}
In the equation above, $(\vec{\omega}^{\prime},-\vec{\omega}^{\prime})$
and $(\vec{\omega} ,-\vec{\omega})$ are sector indices of incoming
and outgoing electron lines respectively.Other sector indices correspond
to the internal electron lines. $\beta$ is a strictly positive number
coming from the fermion loop integration in the  Feynman diagram and is
given by
$ ~\beta =\int dk_{\perp}dk_{\parallel}dk_{0}
 ~[k_{0}^{2} +(v_F k_{\parallel}-\lambda \delta\mu_{1})^{2}]^{-1} ~$.
We find that the renormalized value of $g_{BCS}$ is
\begin{eqnarray}
g_{BCS}^{(j+1)}(\vec{\omega} , \vec{\omega}^{\prime})
&=&g_{BCS}^{(j)}(\vec{\omega} , \vec{\omega}^{\prime})
+\sum_{n=1}^{\infty}\big( \frac{1}{\lambda^{d-1}} \big)^{n}
\sum_{\vec{\omega_1},...,\vec{\omega_n}}
(-1)^n \beta_{j}^{n} g_{BCS}^{(j)}(\vec{\omega},\vec{\omega}_n)
g_{BCS}^{(j)}(\vec{\omega}_n,\vec{\omega}_{n-1}) ...
g_{BCS}^{(j)}(\vec{\omega}_1,\vec{\omega}^{\prime}) \nonumber\\
& & +O(\frac{g^{(j)}}{\lambda_j})
\end{eqnarray}
The explicit expression for flow equation can be obtained by
expanding  the coupling functions
$g_{BCS}(\vec{\omega},\vec{\omega}^{\prime})
=g_{BCS}(\langle \vec{\omega},\vec{\omega}^{\prime})$
into spherical harmonics,
$g_{BCS}(\langle\vec{\omega},\vec{\omega}^{\prime})
=\sum g_{l}h_{l}(\langle\vec{\omega},\vec{\omega}^{\prime})$.
Up to terms of order $\frac{1}{\lambda}$,
\begin{eqnarray}
g_{l}^{(j+1)}= \frac{g_{l}^{(j)}}{1+\beta_{j}g_{l}^{(j)}}
+O\big(\frac{1}{\lambda}\big)
\end{eqnarray}
To obtain the differential equation for the R.G. flow, let us define
$g_{l}^{(j)}$ := $g_{l}^{(\lambda_j)}$, and consider a scale
$\lambda=e^{t}\lambda_0$. Next,
define $g_{l}(t)$:=$g_{l}^{(e^t \lambda_{0})}$. The coefficient
$\beta=\beta(t,t')$ vanishes in the limit $t' \searrow t$, therefore,
$\beta(t',t)=(t'-t)\gamma(t) +O\big( (t'-t)^{2}\big)$. Writing difference
equation for the couplings $g_{l}^{(j+1)}$ and $g_{l}^{(j)}$, and
dividing both sides of the difference equation by $(t'-t)$,
we finally obtain
\begin{eqnarray}
\frac{d}{dt}g_{l}(t)=-\gamma g_{l}(t)(t)^{2} +
O\big(e^{-t}g(t)^{2}\big)
\end{eqnarray}
where $ ~\gamma=\gamma(t)>0 ~$.
It is independent of $l$ and approximately
independent of $t ~$, and therefore, we set $\gamma=\gamma_{0}$.
The positivity of $\gamma$ follows from slow monotone growth of
$\beta (t'-t)$ in $t'$.
Neglecting the error term, the solution can be written as,
\begin{eqnarray}
g_{l}(t)=\frac{g_{l}(0)}{1+\gamma_{0}g_{l}(0)t}.
\end{eqnarray}
If there is an angular momentum channel,$ ~l ~$, with attractive
interactions ($g_{l}(0)<0$) the flow diverges at a finite value
of the scaling parameter, $t=-(\gamma_{0}g_{l}(0))^{-1}$ .
This singularity reflects the BCS-instability
of the ground state.
\section{Instability of the Perturbative Ground State in QED}
We have seen that in the case of large flavour QED as well as the
weakly interacting Fermi system with attractive
potential in any one of the angular momentum channel,
the running coupling constant blow up at some finite value
of their arguments.In the latter case, we know that the
perturbative ground state is unstable (BCS instability):
perturbation theory is built around the wrong ground state
(the non-interacting Fermi gas).
The elementary excitations are the electrons and holes,
and Fock space is constructed using these single particle states.
The BCS instability implies that such excitations
do not exist- we require a modification of the excitation
spectrum.But this also means that the new stable
ground state is not in the eigenstate of particle number
operator, $\hat{N}$ and the global $U(1)$- gauge symmetry
associated with the conservation of particle number (excitations)
is broken. The BCS ground state of superconductivity \cite{froh}
contains all these features.
\par Thus the BCS theory of superconductivity contains two parts :
\par 1)The Landau-Fermi liquid ground state is unstable
\par 2) There exists another ground state- superconducting state
with Cooper pairs as the elementary modes.
\par In this paper, we have been concerned with the first part of the
problem. We have seen that , up to leading order in $1/N$, interactions
in the various angular momentum channels decouple, and attractive
interaction in any angular momentum channel leads to the instability
of (Landau-Fermi liquid) ground state.This analysis in the Feldman
theory was carried out using renormalization group technique in the
language of "Second Quantization". On the other hand the Lieb theory
of stability of matter is discussed in language of
"First Quantization". Both the methods address the issue of stability
of  "Ground State of a Many-Body System". To understand the connection
between the two methods, let us consider a realistic Fermi system. The
most interesting realistic Fermi system happens to be the liquid
$He^{3}$ \cite{woll}.
The minimum of the interatomic potential in $He^{3}$
(as well as in $He^{4}$)  is too close
to the atoms and is too shallow and, therefore the zero point is high
compared to the potential energy and the helium atoms do not bind to form
molecules.However, at higher density (at high pressure) the net potential
energy decreases as $-N^{2}v$ ($v$ being the two body potential) and
overcomes the zero point energy which increases as $N^{5/3}$ with
the particle number $N$.
Thus at higher density, we have the atomic
helium liquid which is a Landau-Fermi liquid.
As the density is increased further potential energy
further decreases.At very high density the zero point energy
is completely overcome,
however the repulsive hard core (two body) potential
comes into play.In $He^{3}$, in the angular mometum channel $l=0$,
the repulson is too strong.However, in $l=1$ and $l=2$ channel the
attractive potential energy over takes the repulson \cite{woll,and1,and2}.
Thus the Lieb theory leads to 'unbounded from bellow' energy in
the two channels (interaction in different angular momentum channel
decouple).This makes the quantum mechanical
Landau-Fermi liquid ground state unstable.
On the other hand, Feldman theory implies that for
the angular momentum channels $l=1$ and $l=2$,
the effective running coupling constants blows up.
\par In the large flavour limit, the effective coupling constant
in QED blows up exactly like the effective  coupling constant
in the Feldman theory. The Lieb treatment of the Coulomb system
also suggests that in the relativistic limit for some value
of the number of flavour $N_f$, the ground state become unstable.
These similarities clearly suggest that the perturbative vacuum state of
QED is unstable. Note that
in the Lieb theory for coulomb systems, we had nonrelativistic
treatment followed by a relativistic system for the stability
of the ground state.In condensed matter systems, the formation of
cooper pairs changes the situation drastically.
The liquid makes a transition to superfluid phase.The cooper pairs
are essentionally free and therefore, the new superfluid state does not
suffer from the type of instability mentioned above.Does the instability
of ground state in QED imply the existense of BCS type of ground state?
The answer seems to be negative.The residue at the poles
of the effective running coupling constant in QED and
in the condensed matter
have opposite signs.In the condensed matter system
it signifies the presence of cooper pair while in the case of
QED the opposite sign signifies the absence
of such pairs \cite{shirkov,abrik}.
\par The remarkable analogy of the evolution equations of the effective
coupling constants in large flavour QED ( Eq.(2) ) and
the condensed matter system above ( Eq.(20) ), lead to the
suggestion that the perturbative ground state (vacuum state)
is unstable in large flavour QED too.
Landau singularity, just as in case of the condensed matter model,
reflects the fact that we have constructed the perturbative theory
around the "wrong" unstable ground state.
In our model of QED, the elementary excitations around the perturbative
vacuum or ground state are the electrons, positrons
and photons and, therefore, instability of ground state or
the vacuum implies that such excitations have become unstable.
The new stable ground state would require modification
of the particle spectrum. There are suggestions in the literature
that the electric charge charge is totally screened by the new
nonperturbative ground state  \cite{landau,kogut1,kogut2,gock,kogut} .
The most interesting and exciting possibility is the one suggested
by Witten \cite{witten}.His conjecture is that at
very high energy the states
carrying electric charge become very massive and are suppressed
while magnetic monopoles become lighter and are excited.

\par Acknowledgement: I would like to thank Prof.J.Froehlich and
Prof.G.M.Graf for hospitality at the Institute of Theoretical
Physics, ETH- Zurich, Switzerland where part of the work was carried
out.


\end{document}